\documentclass[letter]{emulateapj}
\usepackage{natbib}
\usepackage{amsmath}
\usepackage{hyperref}
\usepackage{arydshln}
\usepackage{multirow}
\usepackage{rotating}
\usepackage{threeparttable}

\begin{document}

\title{A Giant Sample of Giant Pulses from the Crab Pulsar}

\author{M.~B.~Mickaliger\altaffilmark{1}, M.~A.~McLaughlin\altaffilmark{1,2}, D.~R.~Lorimer\altaffilmark{1,2}, G.~I.~Langston\altaffilmark{3}, A.~V.~Bilous\altaffilmark{4}, V.~I.~Kondratiev\altaffilmark{5,6}, M.~Lyutikov\altaffilmark{7}, S.~M.~Ransom\altaffilmark{8}, \& N.~Palliyaguru\altaffilmark{1}}

\altaffiltext{1}{Department of Physics, West Virginia University, Morgantown, WV 26506}
\altaffiltext{2}{Also adjunct at the National Radio Astronomy Observatory, Green Bank, WV 24944}
\altaffiltext{3}{National Radio Astronomy Observatory, Green Bank, WV 24944}
\altaffiltext{4}{Department of Astronomy, University of Virginia, PO Box 400325, Charlottesville, VA 22904}
\altaffiltext{5}{Netherlands Institue for Radio Astronomy (ASTRON), Postbus 2, 7990 AA Dwingeloo, The Netherlands}
\altaffiltext{6}{Astro Space Center of the Lebedev Physical Institute, Profsoyuznaya str. 84/32, Moscow 117997, Russia}
\altaffiltext{7}{Department of Physics, Purdue University, 525 Northwestern Avenue, West Lafayette, IN 47907-2036}
\altaffiltext{8}{National Radio Astronomy Observatory, Charlottesville, VA 22903}

\begin{abstract} 

We observed the Crab pulsar with the 43-m telescope in Green Bank, WV over a timespan of 15 months. In total we obtained 
100 hours of data at 1.2 GHz and seven hours at 330 MHz, resulting in a sample of about 95000 giant pulses (GPs). This is 
the largest sample, to date, of GPs from the Crab pulsar taken with the same telescope and backend and analyzed as one data 
set. We calculated power-law fits to amplitude distributions for main pulse (MP) and interpulse (IP) GPs, resulting in 
indices in the range of 2.1$-$3.1 for MP GPs at 1.2 GHz and in the range of 2.5$-$3.0 and 2.4$-$3.1 for MP and IP GPs at 
330 MHz. We also correlated the GPs at 1.2 GHz with GPs from the Robert C. Byrd Green Bank Telescope (GBT), which were 
obtained simultaneously at a higher frequency (8.9 GHz) over a span of 26 hours. In total, 7933 GPs from the 43-m telescope 
at 1.2 GHz and 39900 GPs from the GBT were recorded during these contemporaneous observations. At 1.2 GHz, 236 (3\%) MP GPs 
and 23 (5\%) IP GPs were detected at 8.9 GHz, both with zero chance probability. Another 15 (4\%) low-frequency IP GPs were 
detected within one spin period of high-frequency IP GPs, with a chance probability of 9\%. This indicates that the 
emission processes at high and low radio frequencies are related, despite significant pulse profile shape differences. The 
43-m GPs were also correlated with \emph{Fermi} $\gamma$-ray photons to see if increased pair production in the 
magnetosphere is the mechanism responsible for GP emission. A total of 92022 GPs and 393 $\gamma$-ray photons were used in 
this correlation analysis. No significant correlations were found between GPs and $\gamma$-ray photons. This indicates that 
increased pair production in the magnetosphere is likely not the dominant cause of GPs. Possible methods of GP production 
may be increased coherence of synchrotron emission or changes in beaming direction.

\end{abstract}


\section{Introduction}

The Crab pulsar was discovered by Staelin \& Reifenstein in 1968 through its giant pulses. Giant pulses (GPs) can be 
thousands of times brighter than the average pulse. The temporal occurence of GPs is random but at frequencies below 3 GHz 
they always occur at the phase of either the Crab pulsar's main pulse (MP) or interpulse (IP) \citep{Lundgren:1994}. At 
frequencies from 4 GHz to 8.4 GHz, GPs are emitted at the phases of the MP and IP as well as at the phases of two additional 
high-frequency components \citep{Moffett:1996}. Above 8.5 GHz, GPs are again seen at only the phases of the MP and IP 
\citep{Jessner:2010}

\begin{figure}[ht]
\centering
\includegraphics[scale=0.65, angle=-90]{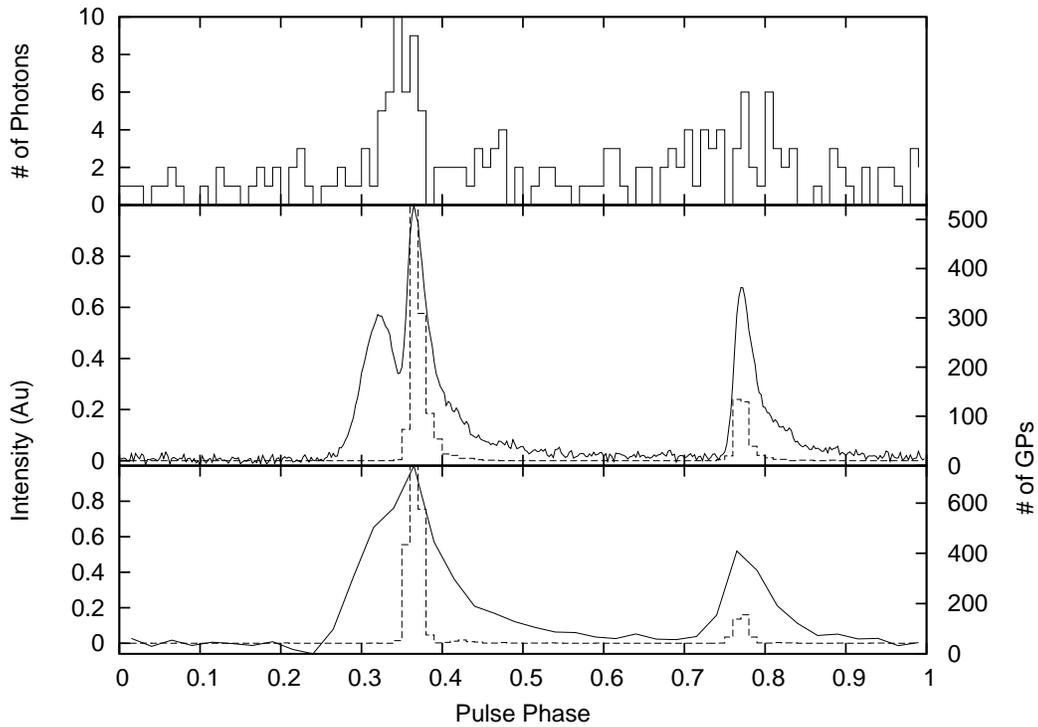}
\caption{Folded pulse profiles from MJD 55406 for \emph{Fermi} (top), the GBT (middle, solid line), and the GB43 (bottom,
solid line), as well as histograms of the number of GPs for the GBT (middle, dashed line) and GB43 (bottom, dashed line).
The \emph{Fermi} profile is made using photons above 100 MeV over a 24 hour period ($\sim$11 hours on source time). The GB43
observations were taken at a center frequency of 330 MHz over a 220 MHz band for 3.7 hours and the GBT observations were
taken at a center frequency of 350 MHz over a 100 MHz band for 10 minutes. The folded radio profiles have been normalized
to have peaks of unity.}
\label{fig:all_profiles}
\end{figure}

The emission mechanism of GPs is still an open question \citep[e.g.][]{Petrova:2006, Istomin:2004, Weatherall:1998}. GPs 
could be caused by increased pair production in the magnetosphere, increased coherence of synchrotron emission, or changes 
in beaming direction. Correlating radio GPs with high-energy photons is one way to determine if increased pair production 
is a major cause of GPs. A recent model \citep{Lyutikov:2007} proposes that GPs are generated on the last closed magnetic 
field line near the light cylinder via anomalous cyclotron resonance. If this is true, there would be an increase in 
$\gamma$-rays at the times of radio GPs. Anomalous cyclotron resonances have been previously proposed as a cause of GP 
emission by \citet{Lyutikov:1999} and \citet{Machabeli:1979}. We expect the $\gamma$-rays to be phase aligned with the 
radio GPs at 1.2 GHz as the $\gamma$-ray and 1.2 GHz radio profiles are aligned (see Figure \ref{fig:all_profiles}). Since 
we have such a large data set of GPs, we can correlate them with $\gamma$-ray photons to test this model. Correlations 
between radio GPs and $\gamma$-ray photons have been carried out previously by \citet{Bilous:2011} and 
\citet{Lundgren:1995}. \citet{Shearer:2003} correlated radio GPs and optical photons and found a slight correlation, and 
work by \citet{Collins:2012} supports this result.

The \emph{Fermi} Large Area Telescope (LAT) is a pair conversion telescope that operates in the energy range from 20 MeV to 
300 GeV. It has a large field of view (2.4 sr), which allows it to rapidly map the entire sky, and very good angular 
resolution, minimizing background contamination \citep{Atwood:2009}. The Crab pulsar is a bright \emph{Fermi} source and 
has been studied at these energies. \citet{Abdo:2010} found that the $\gamma$-ray profile is double peaked and matches the 
1.4 GHz profile, with the $\gamma$-ray peaks leading the radio peaks by $\sim$0.01 phase. Since the LAT covers the energies 
predicted by \citet{Lyutikov:2007} and observes the Crab pulsar multiple times per day, we can use \emph{Fermi} data to 
test Lyutikov's theory.

The power-law nature of the amplitude distribution of GPs is well known \citep[e.g.][]{Argyle:1972, Popov:2007}, but there 
have been varying values for the power-law index calculated, even for similar frequencies \citep[e.g.][]{Karuppusamy:2010, 
Bhat:2008}. With our large sample of GPs, we can calculate power-law indices over a long timespan. Since the amplitude 
distribution of pulses from many pulsars is log-normal \citep{Ritchings:1976}, GPs must have a different emission 
mechanism. By comparing our power-law index with those of other neutron star source classes (pulsars, magnetars, RRATs), we 
may be able to put constraints on the GP emission mechanism.

A $\gamma$-ray flare was recently observed from the Crab Nebula by the AGILE satellite \citep{Tavani:2011}. The flare 
lasted from MJDs 55457$-$55461 and was also observed by \emph{Fermi} \citep{Hays:2010}. This occured during the span of our 
observations, so we can compare radio and $\gamma$-ray properties of the pulsar from before the flare with a few days of 
data taken about two months after the flare to verify that the flare was not associated with the pulsar. Current theories 
suggest that the flare was caused by electrons accelerated by magnetic reconnection at the termination shock of the Crab 
Nebula \citep[e.g.][]{Bednarek:2011, Bykov:2012, Cerutti:2012}.

The plan for the rest of this paper is as follows. Section \ref{radobs} describes the radio observations and outlines the 
radio analysis. Section \ref{stats} discusses the GP statistics and the power-law index calculations. Section \ref{gbtcorr} 
details the correlation between 43-m and GBT GPs. Section \ref{fermicorr} presents the \emph{Fermi} data used and its 
reduction as well as the 43-m/\emph{Fermi} correlation. Section \ref{flare} discusses the recent $\gamma$-ray flare. 
Finally, conclusions are offered in Section \ref{conc}.

\section{Radio Observations}
\label{radobs}

We collected data over the course of 15 months using the 43-m telescope (hereafter GB43), located at the National Radio 
Astronomy Observatory site in Green Bank, WV. The GB43 is funded by MIT Lincoln Labs to perform bistatic radar observations 
of the ionosphere \citep{Langston:2007}. Most of the observations had a center frequency of 1.2 GHz with a useable 
bandwidth of 400 MHz, with a handful of the later observations centered at 330 MHz, with a useable bandwidth of 150 MHz. 
All of the observations were taken using 4096 frequency channels and varying sampling times, listed in Table 
\ref{table:crab_data}. Some of the sampling times used were rather long, and this was done to decrease the data volume. In 
total, we observed the Crab pulsar for 100 hours at 1.2 GHz and seven hours at 330 MHz. The 330 MHz mode was implemented 
later and used because pulsars, in general, are stronger at lower frequencies and exhibit a steep spectral index 
\citep{Lorimer:1995}. Even though the smaller bandwidth at 330 MHz reduces sensitivity by almost a factor of 2, the steep 
spectrum of the Crab pulsar ($\alpha$~=~3.1$\pm$0.2 \citep{Lorimer:1995}, where S~=~$\nu^{-\alpha}$) means that the pulsar 
is 55 times brighter at 330 MHz than at 1.2 GHz. The flux density from the Crab Nebula is greater at lower frequencies 
\citep{Cordes:2004} but, due to a shallow power-law index ($\nu^{-0.27}$) \citep{Allen:1973, Bietenholz:1997}, at 330 MHz 
(flux density $\sim$1288~Jy) is only 1.4 times brighter than at 1.2 GHz (flux density $\sim$909~Jy). Therefore there should 
be many more GPs above our 10$\sigma$ threshold at 330 MHz than at 1.2 GHz.

Data were taken with WUPPI, the West Virginia University Ultimate Pulsar Processing Instrument. WUPPI is a clone of GUPPI 
(Green Bank Ultimate Pulsar Processing Instrument) \citep{DuPlain:2008} for use with the GB43. GUPPI is a flexible digital 
backend for the GBT. Like GUPPI, WUPPI is built from reconfigurable off-the-shelf hardware and software available from the 
CASPER (Center for Astronomy Signal Processing and Electronics Research) group \citep{Parsons:2009}. Both GUPPI and WUPPI 
sample the data with 8-bit precision over bandwidths as large as 800 MHz, and are capable of recording all four Stokes 
parameters. To ease disk space usage, only total intensity (Stokes I) was recorded.

The data were processed using a real-time data reduction pipeline on a 16 processor mini-cluster. The pipeline is comprised 
of a set of scripts, built from freely available analysis software\footnote{\url{http://sigproc.sourceforge.net}}, that can 
reduce a file in real time. First, a file in PSRFITS format \citep{Hotan:2004} has its mean bandpass divided out. 
Frequency channels that have intensities above the resulting mean are flagged as containing radio frequency interference 
(RFI) and removed. Then the resulting file is dedispersed at both a dispersion measure (DM) of zero and the DM of the Crab, 
which is $\sim$56.8~pc cm$^{-3}$; the exact values for each observation, obtained from the Jodrell Bank Crab Pulsar Monthly 
Ephemeris\footnote{\url{http://www.jb.man.ac.uk/~pulsar/crab.html}} \citep{Lyne:1993}, are listed in Table 
\ref{table:crab_data}. The DM of the Crab pulsar can vary on a monthly timescale by $\sim$0.01~pc cm$^{-3}$. In order to 
keep processing in real-time we did not correct for this. To ease space requirements, raw data were not kept, so we could 
also not correct for this later. Errors in GP arrival times due to an incorrect DM on the order of 0.01~pc cm$^{-3}$ are 
$\sim$20~$\mu$s, which is less than the sampling time. Both resulting time series are then 
searched for single pulses with a peak signal-to-noise (S/N) above 10$\sigma$ using our single pulse search (based on the 
algorithm described by \citet{Cordes:2003}). Our 10$\sigma$ definition of a GP is somewhat arbitrary because there is no 
set threshold for a GP as the weakest GPs have yet to be observed. The current population of observed GPs accounts for no 
less than 50\% of the pulsed emission at frequencies around 1.6 GHz, and the inclusion of the weakest GPs could bring that 
number up to 90\% \citep{Majid:2011}. A GP time-of-arrival (TOA) (measured by taking the arrival time of the peak of the 
pulse), the S/N of that peak, and the pulse width, taken as the width of a best-fit top hat function, are recorded for 
each pulse. The TOAs from each time series are compared and if a pulse detected at the DM of the Crab is within 0.1 s of a 
zero DM pulse, the S/N is checked for both pulses\footnote{This is of the order of the frequency dependent arrival time 
delay across the band, and is used as it removes very nearly all RFI, while removing few real GPs.}. If the S/N is higher 
in the zero DM pulse, the pulse is assumed to be due to RFI and removed. The pipeline outputs a profile for each GP and 
produces an average folded profile for the entire observation, a plot of GP arrival time vs pulse phase, and an average GP 
profile, made by summing all of the individual GP profiles.

\begin{figure}[ht]
\centering
\includegraphics[scale=0.65, angle=-90]{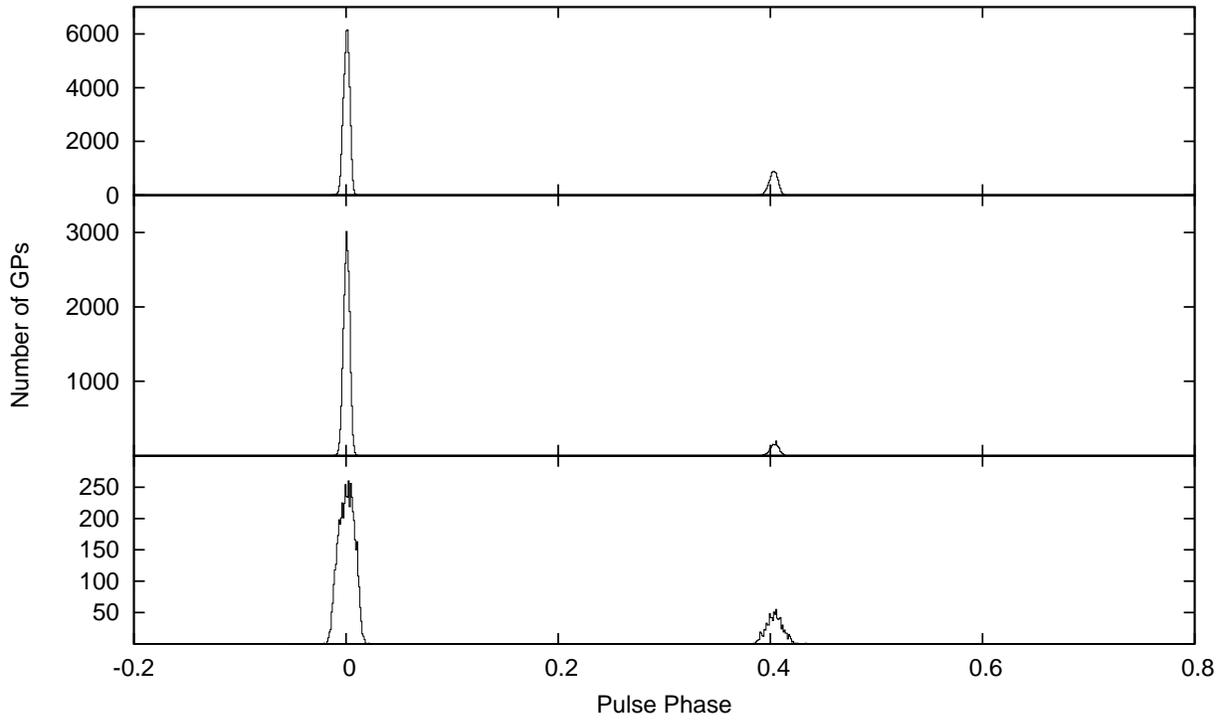}
\caption{Number of GPs versus pulse phase for all GPs collected with the GB43 at 1.2 GHz (top, post-flare; middle, 
pre-flare) and 330 MHz (bottom). Most GPs at frequencies below 5 GHz come at the phase ($\sim$0) of the MP, while there are 
still a considerable number at the IP phase ($\sim$0.4). No GPs are seen at other phases. The FWHMs of the 1.2 GHz 
distributions (226~$\mu$s pre-flare, 292~$\mu$s post-flare) are on the order of the FWHMs of the folded profiles at 1.2 
GHz, while the FWHM of the 330 MHz distribution (704~$\mu$s) is much narrower than the folded profile and is on the order 
of the average GP profile at 330 MHz. Even though there were only four observing epochs after the flare, there were many 
more GPs recorded than pre-flare due to a large increase in receiver sensitivity.}
\label{fig:pulse_hist}
\end{figure}

Figure \ref{fig:pulse_hist} shows the number of GPs detected versus pulse phase. It is worth noting that, although there 
were only four post-flare observing epochs, there are many more GPs than were recorded in all of the pre-flare data. This 
was due to a receiver upgrade which resulted in a large increase in sensitivity between our observations on MJD 55412 and 
MJD 55516. At frequencies below 5 GHz \citep{Cordes:2004}, most detected GPs come at the phase of the MP, with the rest at 
the IP phase. In our data, $\sim$87\% of detected GPs were at the phase of the MP, while the other $\sim$13\% were at the 
phase of the IP. Only 5\% of GPs detected by \citet{Cordes:2004} at 1.2 GHz were at the phase of the IP, but 
\citet{Karuppusamy:2010} found that $\sim$12\% of GPs detected at 1.4 GHz were IP GPs. For detected pulses that show up out 
of phase with the MP and IP, the frequency versus time plots are checked by eye to see if they show the proper quadratic 
frequency sweep for a DM of 56.8~pc cm$^{-3}$. If they do not, those pulses are removed. In most cases, the RFI removal 
process mentioned above removes most of these false detections before they are checked by eye. We found no events at phases 
other than those of the MP or IP that showed the proper frequency sweep to be a real GP.

Most of the GPs we detected have a constant intensity across the entire band. Some of them, however, show variations 
in amplitude as a function of frequency, which have been seen before \citep[e.g.][]{Karuppusamy:2010}. We also see 
variations in amplitude between days, which is likely due to refractive interstellar scintillation (RISS), which can affect 
the strength of pulses on a timescale of days.

\begin{figure}[ht]
\centering
\includegraphics[scale=0.65, angle=-90]{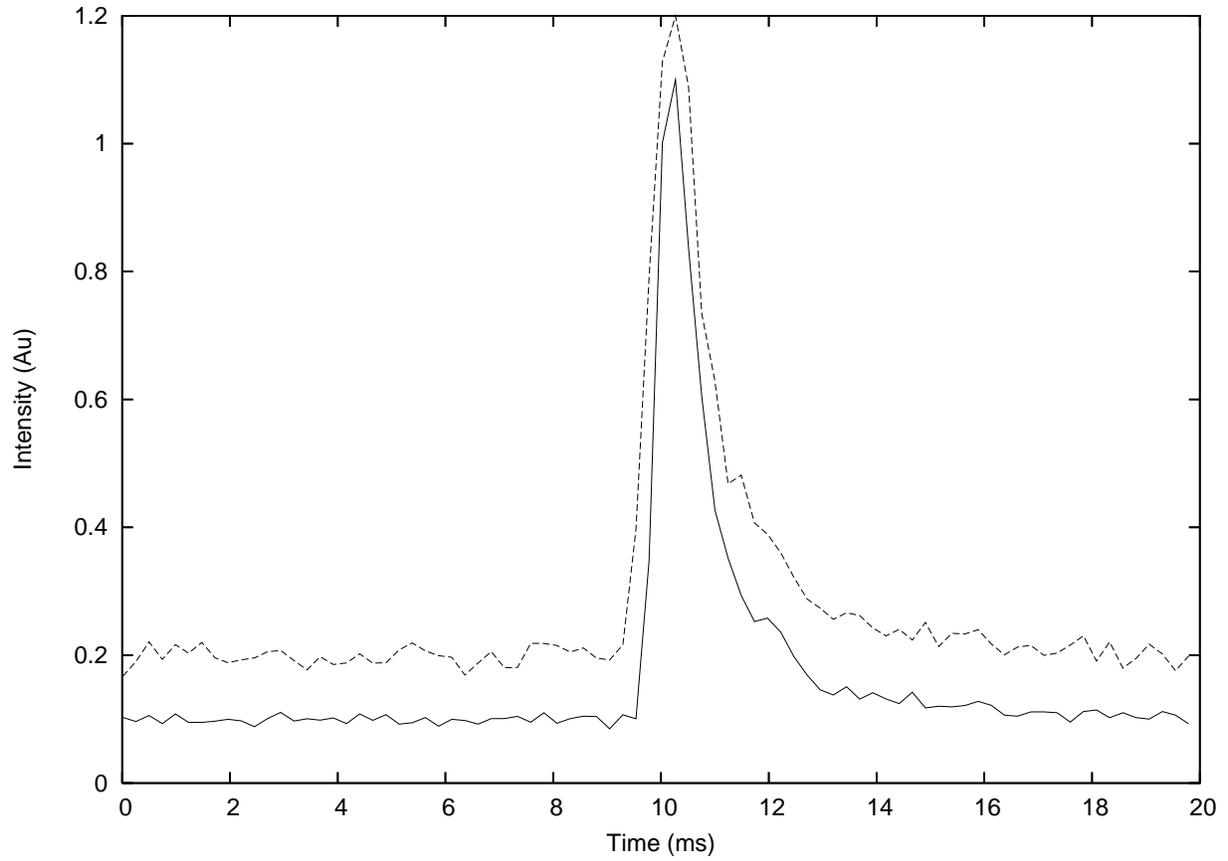}
\caption{A single GP observed both with the GB43 (dashed line) at 330 MHz and the GBT (solid line) at 350 MHz on MJD 55406. 
When corrected for the dispersion delay due to different observing frequencies, the peaks of the GP from both observations 
are within one $\mu$s of each other. The GB43 profile has been shifted upwards for clarity.}
\label{fig:gp_ref}
\end{figure}

In order to determine if the time stamps for the GB43 are accurate, a short ($\sim$10 min) observation of the Crab pulsar 
was made contemporaneously on MJD 55406 at similar frequencies with the GB43 and GBT. The GB43 used a center frequency of 
330 MHz, while the GBT used 350 MHz. GPs were detected in each data set and their TOAs were compared. As can be seen in 
Figure \ref{fig:gp_ref}, the pulse shapes of contemporaneous GPs are similar, and the TOAs, converted to infinite frequency 
at the solar system barycenter, have identical arrival times to within the instrumental resolution.

\begin{figure}[ht]
\centering
\includegraphics[scale=0.65, angle=-90]{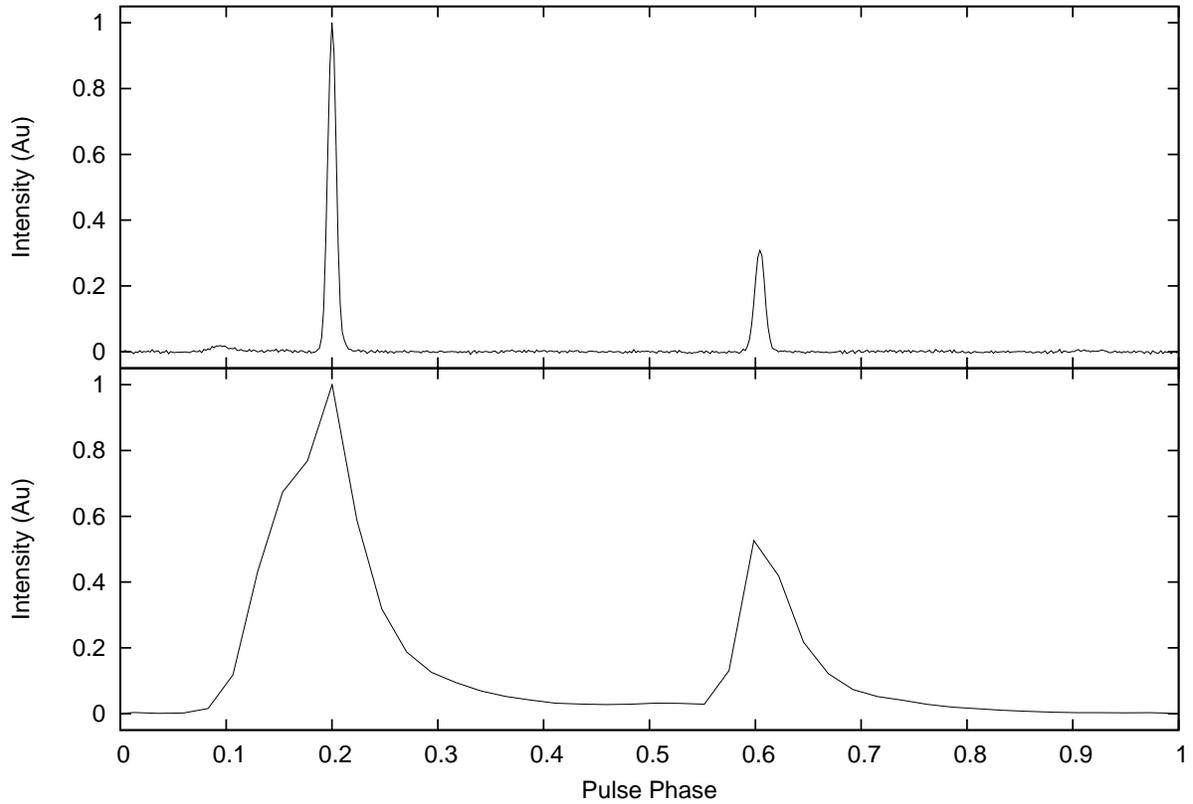}
\caption{Average pulse profiles from all 1.2 GHz (top) and 330 MHz (bottom) observations with the GB43, listed in Table 
\ref{table:crab_data}. The 1.2 GHz profile excludes MJDs 55097, 55257, 55290, 55299, and 55346, where the folded pulse 
profiles were dominated by RFI. At 1.2 GHz the MP (larger) and IP (smaller) are apparent, and a low-frequency component can 
be seen one-tenth of a pulse phase ahead of the MP \citep[e.g.][]{Moffett:1996}. At 330 MHz the MP and IP are much wider 
due to large DM smearing, which at the top of the band (440 MHz) is 0.3~ms and at the bottom (220 MHz) is 2.4~ms. The 
scattering time, estimated from \citet{Kuzmin:2011}, is 470~$\mu$s at the top of the band and 5.8~ms at the bottom, and the 
sampling time is 819.2~$\mu$s. This leads to an effective resolution of 990~$\mu$s at the top of the band and 6.3~ms at the 
bottom.}
\label{fig:av_profiles}
\end{figure}

Figure \ref{fig:av_profiles} shows the average profile of the Crab pulsar at 1.2 GHz and 330 MHz. The intensities are 
in arbitrary units. The MP and IP are visible at both frequencies. The increased width at low-frequency is due largely to 
DM smearing, which is on the order of 2.4~ms at the bottom of the band. The 1.2 GHz profile exhibits a weak low-frequency 
component (LFC), about one-tenth of a pulse phase ahead of the MP \citep[e.g.][]{Moffett:1996}. At 330 MHz there is a 
precursor (PR) to the MP \citep[e.g.][]{Rankin:1970} which is broad enough at this frequency that it shows up as the first 
peak of the double-peaked MP \citep[see][for the frequency evolution of the PR]{Karuppusamy:2012}, which is visible in the 
low-frequency GBT profile from MJD 55406 (Figure \ref{fig:all_profiles}). Due to a much larger effective resolution, the MP 
in the GB43 profile is unresolved and appears to have a single peak. Table \ref{table:times} lists the effective 
resolutions at the top and bottom of the band for both the 330 MHz and 1.2 GHz GB43 observations as well as the GBT 350 MHz 
observation.

\begin{figure}[ht]
\centering
\includegraphics[scale=0.65, angle=-90]{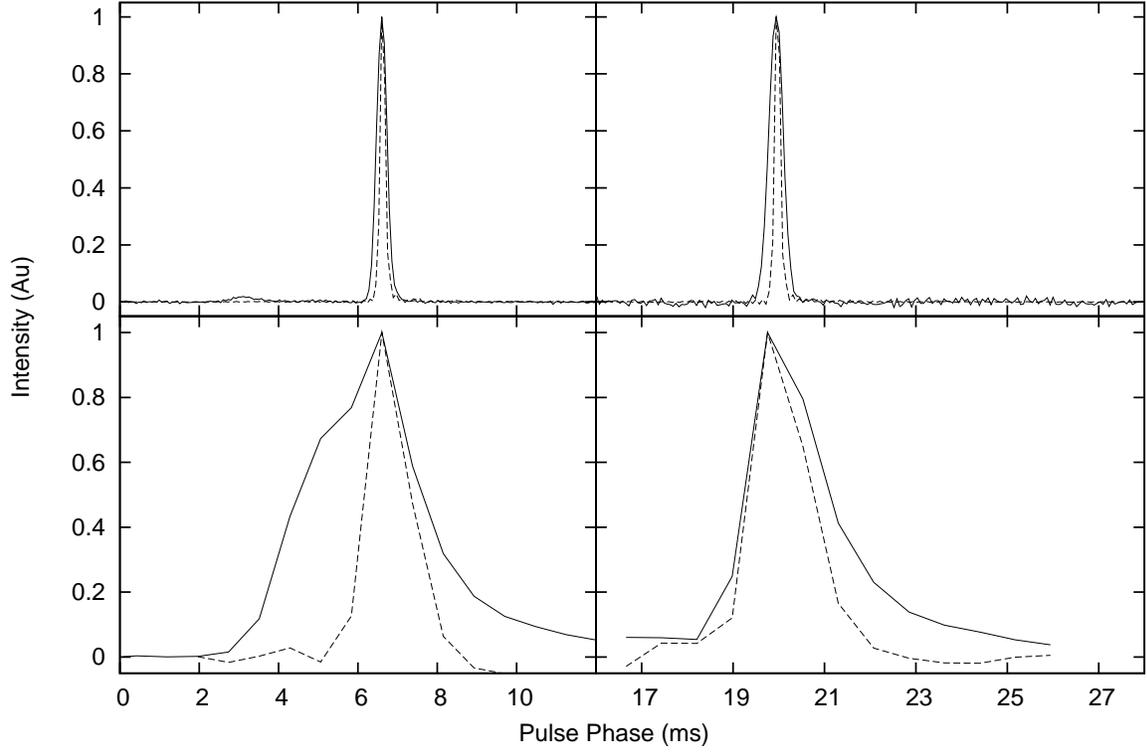}
\caption{Average GP profiles for all MP (left) and IP (right) GPs (dashed lines) collected with the GB43 at 1.2 GHz (top) 
and 330 MHz (bottom). The 1.2 GHz profiles are made by summing the 76707 individual MP and 10871 individual IP GP profiles, 
and the 330 MHz profiles are made by summing the 5232 individual MP and 888 individual IP GP profiles. They are shown with 
the folded MP and IP profiles for all 1.2 GHz and 330 MHz data (solid lines). The FWHM of the 1.2 GHz folded MP and IP are 
$\sim$305~$\mu$s and $\sim$360~$\mu$s, respectively, while the FWHM of the average MP and IP GPs are both $\sim$140~$\mu$s. 
The FWHM of the 330 MHz folded MP and IP are $\sim$3.2~ms and $\sim$1.5~ms, respectively, while the FWHM of the average MP 
and IP GPs are $\sim$1.1~ms. The 330 MHz MP is much wider than the IP due to its overlap with the MP precursor. The 
intensities of the profiles are arbitrary and are scaled for clarity.}
\label{fig:prof_comp}
\end{figure}

The full-widths-at-half-maximum (FWHMs) of the MP and IP from the GB43 profiles are $\sim$305~$\mu$s and $\sim$360~$\mu$s 
at 1.2 GHz and $\sim$3.2~ms and $\sim$1.5~ms at 330 MHz. At 330 MHz, the MP is wider than the IP due to its overlap with 
the PR \citep{Rankin:1970}. The average GP profiles, however, are narrower than the average folded profiles at both 
frequencies (see Figure \ref{fig:prof_comp}), with FWHMs of $\sim$140~$\mu$s for both the MP and IP at 1.2 GHz, and 
$\sim$1.1~ms at 330 MHz for both the MP and IP. It has been suggested by \citet{Popov:2006} 
that every pulse at the phase of the MP and IP is a GP, and that normal emission only comes from the PR, where no GPs have 
been seen. We also have not seen any GPs at the phase of the PR at 330 MHz in our data. Even though our observed GP 
profiles are narrower than the average folded profiles, this hypothesis is not ruled out. \citet{Popov:2007} found that 
pulse width is inversely proportional to pulse intensity, so lowering our S/N definition of a GP would include wider GPs 
which could possibly increase the width of the average GP profile to that of the folded profile. Also, the phases of weaker 
GPs may have more deviation from the center phase of the MP and IP than stronger GPs, leading to a wider average profile. 
Unfortunately, we were unable to lower our GP threshold below 10$\sigma$ as this resulted in many spurious pulses.

\section{Amplitude Distributions}
\label{stats}

A total of 93698 GPs were observed with the GB43. At 1.2 GHz we recorded 76707 GPs at the phase of the MP and 10871 GPs at 
the phase of the IP, and at 330 MHz 5232 MP and 888 IP GPs were recorded. Due to low GP statistics on some days, only 
78574 MP GPs and 9693 IP GPs were used in fitting the amplitude distributions.

\begin{figure}[ht]
\centering
\includegraphics[scale=0.65, angle=-90]{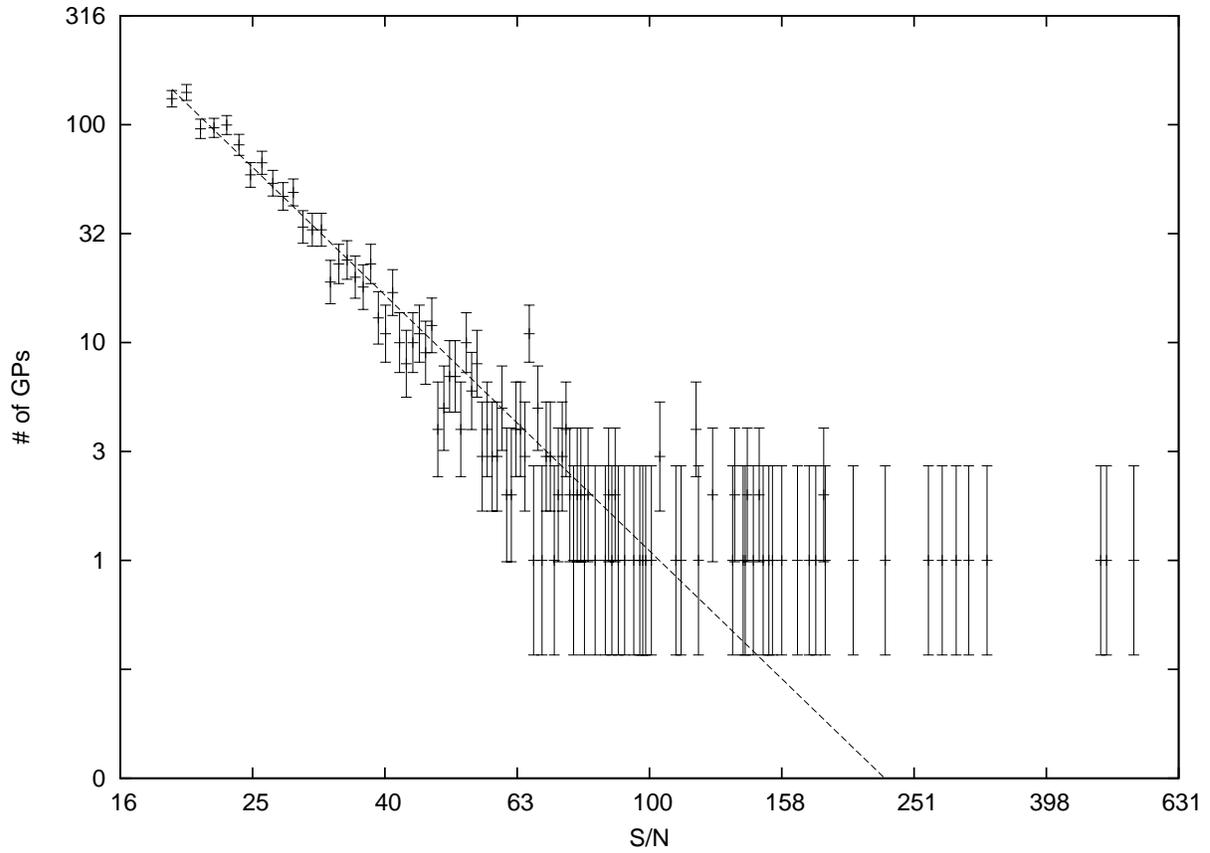}
\caption{A plot of the log of the number of GPs collected with the GB43 vs the log of their S/N for the MP at 1.2 GHz on 
MJD 55099 and a best-fit power-law, which has a slope of 2.93$\pm$0.07. The deviation from the power-law distribution at 
high S/N is seen on all days, and these GPs may be examples of supergiant pulses, seen previously by \citet{Cordes:2004}. 
The average supergiant pulse is about 15 times stronger than the average GP.}
\label{fig:amp_dist}
\end{figure}

We calculated, through least-squares fitting, power-law indices for the differential amplitude distributions of MP and IP 
GPs at both 1.2 GHz and 330 MHz for each day separately. Figure \ref{fig:amp_dist} is an example of a power-law fit for the 
MP. We noted that the power-law index of GPs varied daily, as was previously reported by \citet{Lundgren:1995} at similar 
frequencies. This gave a range of power-law indices from 2.1$\pm$0.3 to 3.1$\pm$0.2 for MP GPs at 1.2 GHz, 2.4$\pm$0.4 to 
2.81$\pm$0.03 for IP GPs at 1.2 GHz, 2.5$\pm$0.2 to 2.95$\pm$0.09 for MP GPs at 330 MHz, and 2.4$\pm$0.2 to 3.1$\pm$0.2 for 
IP GPs at 330 MHz. Table \ref{table:indices} lists the power-law indices for each day. These indices agree with other 
published power-law indices, listed in Table \ref{table:indices_comp}. The range of indices is likely due to wider GPs 
having steeper spectra, as was seen by \citet{Popov:2008}. We did not have the time resolution to separate the GPs by 
width, as the widest GPs reported by \citet{Popov:2008} were 64~$\mu$s, which is the shortest sampling time we used.

We were unable to properly calibrate the data due to lack of any off-source pointings and/or observations with a pulsed 
cal. There were also doubts as to the stability of the receiver over long timespans. We were able to do a rough 
calibration, however, by correcting for RISS, which affects the strength of all GPs by the same amount, so there should be 
no change to the power-law slope. \citet{Rickett:1990} found that the RISS timescale for the Crab pulsar scales as 
$\nu^{-2.2}$. Based on this, the timescales at 1.2 GHz and 330 MHz are 1.7 days and 30 days, respectively. A Lomb-Scargle 
analysis \citep{Scargle:1982} of the periodicity of GP arrival times yielded periodicities of 0.41 days at 330 MHz and 0.99 
days at 1.2 GHz. The $\sim$one day periodicity of the GP arrival times in the 1.2 GHz data is on the order of the RISS 
timescale at that frequency, but the same periodicities were also seen in the randomized data set, so they are not 
significant. We attempted to correct for RISS by scaling observations based on the brightness of their folded profiles. 
However, if GPs dominate the average folded profile, as was seen by \citet{Popov:2006}, any intrinsic variations in their 
strength will affect this correction. To determine if this was true, we removed the GPs from the average profile. We found 
that at 1.2 GHz the amplitude of the average profile without GPs was 1\% less than the average profile with the GPs, and at 
330 MHz was 2\% less. We therefore concluded that GPs do not dominate the average profile. One possible reason why our 
results are the opposite of what was seen by \citet{Popov:2006} is that they recorded all GPs above 5$\sigma$, while we 
were only able to record GPs above 10$\sigma$. Therefore, it is possible that weaker GPs dominate the average profile. If 
the average profile is truly dominated by GPs, then correcting for RISS is complicated, but since we do not see this, we 
used the following correction.

We took the brightest folded profile from each of our three observing epochs (pre-flare 1.2 GHz, 330 MHz, 
post-flare 1.2 GHz) and scaled all of the folded profiles from those epochs to that profile. That provided us with a 
scaling factor for each day, by which we then multiplied the GP S/Ns. We also compared our indices with power-law indices 
of other source classes in an attempt to constrain the GP emission physics. Other source classes and their indices are 
listed in Table \ref{table:source_class}. Although most normal pulsars have log-normal distributions, some pulsars have 
amplitude distributions that have power-law tails. These power-law exponents are included in Table \ref{table:source_class} 
under `Normal Pulsars'. Our power-law indices match those of magnetars and RRATs, but do not match the power-law tails seen 
from normal pulsars.

Also seen in Figure \ref{fig:amp_dist} is a non power-law tail, which is seen on all days in both the MP and IP. This 
deviation is significant, and was also seen by \citet{Cordes:2004}, who postulated these outlying GPs could be supergiant 
pulses, indicating that there may be two distinct mechanisms for GP generation. These supergiant pulses account for 
slightly less than one percent of the GPs used in the amplitude distributions.

\section{Correlating Radio GPs from the GB43 and GBT}
\label{gbtcorr}

\begin{figure}[ht]
\centering
\includegraphics[scale=0.65, angle=-90]{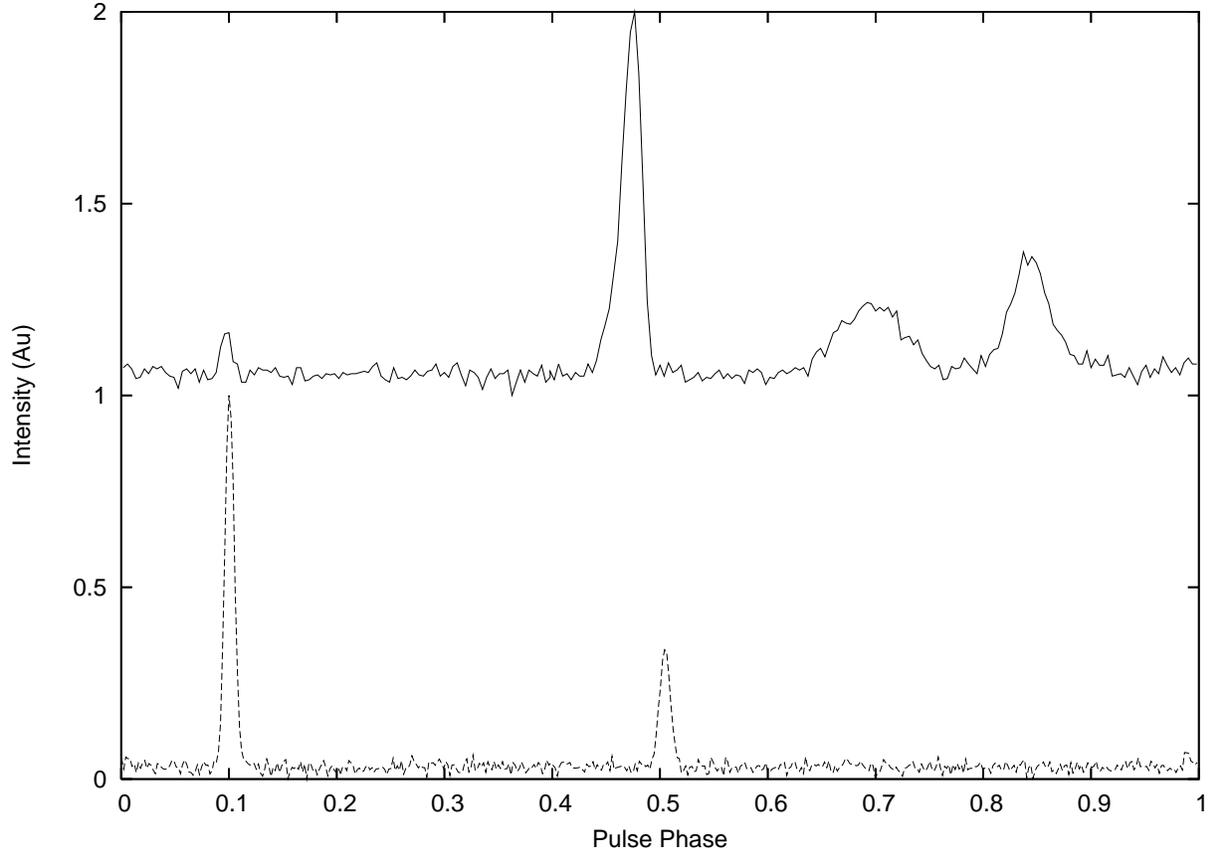}
\caption{Average profiles from the GBT at 8.9 GHz (solid line, from \citet{Bilous:2011}) and the GB43 at 1.2 GHz (dashed
line). The weakening of the MP at high frequencies can be seen, as well as the strengthening and shift of the IP.}
\label{fig:profiles}
\end{figure}

For 16 hours over the span of eight days, we observed the Crab pulsar simultaneously with the GB43 and GBT. The center 
frequency for GB43 observations was 1.2 GHz with 400 MHz of useable bandwidth, while that of the GBT was 8.9 GHz with 800 
MHz of useable bandwidth. At high frequencies the majority of GPs come at the phase of the IP, with fewer MP GPs and few 
GPs from high-frequency components (HFCs) (see Figure \ref{fig:profiles} for a comparison of the folded profile at high and 
low frequencies). For this work, we did not use any HFC GPs in our correlation analysis. We matched the barycentered 
arrival times of the 39900 GPs with S/N $>$ 10 recorded with the GBT (1035 MP GPs, 38865 IP GPs) with the 7933 GPs recorded 
simultaneously with the GB43 (7466 MP GPs, 467 IP GPs). We found that 236 low-frequency MP GPs were also detected 
simultaneously at 8.9 GHz. These GPs were neither the strongest nor the weakest pulses from either data set. The chance 
probability of this occuring is zero percent. This is not surprising, as we know that the high- and low-frequency MP are 
the same component. We did expect, however, that all of the MP GPs detected at 8.9 GHz would be the strongest of the MP GPs 
detected at 1.2 GHz, since the MP is significantly weaker at 8.9 GHz than at 1.2 GHz. All chance probabilities were 
calculated assuming Poisson statistics, using the formula $P=\frac{e^\lambda\times\lambda^K}{K!}$, where $\lambda$ is the 
number of detections expected and $K$ is the number of detections actually recorded.

\begin{figure}[ht]
\centering
\includegraphics[scale=0.65, angle=-90]{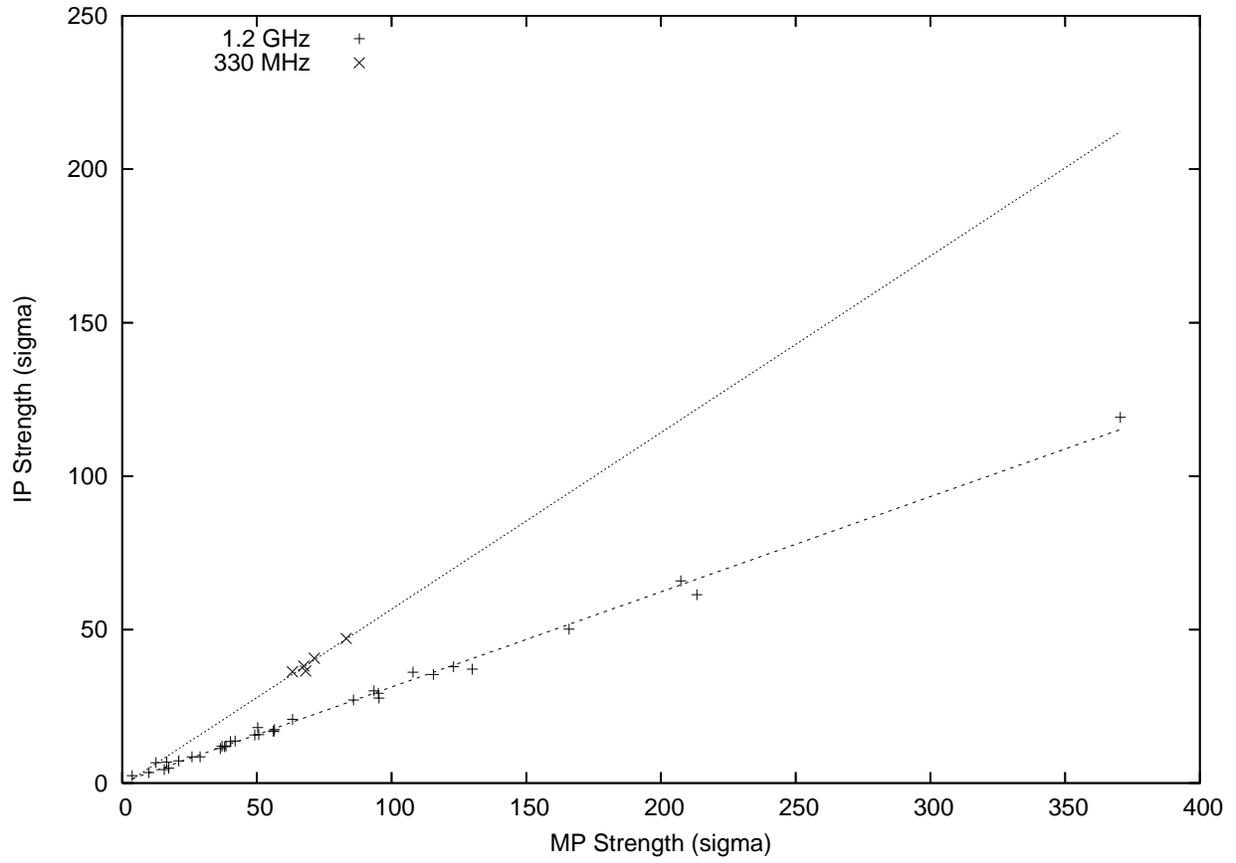}
\caption{Comparison of the strength of the MP of the folded profile to that of the IP for each observation. There is an 
obvious linear trend in both the 330 MHz and 1.2 GHz data, with slopes of 0.58$\pm$0.08 and 0.310$\pm$0.004, respectively. 
This supports the theory that the IP may be a reflection of the MP off of the magnetosphere.}
\label{fig:mpvsip}
\end{figure}

\citet{Moffett:1996} first noticed that the low- and high-frequency IP components of the folded profile are separated by 
10$^{\circ}$, which is about 970~$\mu$s. We checked to see if any low-frequency IP GPs were simultaneously detected at 8.9 
GHz. Only 23 were found ($\sim$5\%), with a zero percent chance occurence, but this is expected since the spectral index of 
GPs is steep \citep{Popov:2008}. We then checked to see if low- and high-frequency IP GPs commonly occured within the same 
rotation period of the pulsar, as was previously seen by \citet{Popov:2008}. In our data, we found 15 instances ($\sim$4\%) 
of low-frequency IP GPs occuring within one spin period of a high-frequency IP GP. The probability of this measurement 
happening by chance is 9\%. In four of these instances the GPs are within one $\mu$s of each other and the chance 
probability of this occuring is zero. In the other 11 instances, the GPs are almost one spin period apart. These are likely 
statistical, as there is a 10\% chance of this occuring randomly. The fact that only four of the 1.2 GHz IP GPs occur 
within one $\mu$s of an 8.9 GHz IP GP suggests that the high and low-frequency IP may be created by different physical 
processes \citep{Moffett:1997} possibly due to emission from different regions in the magnetosphere \citep{Hankins:2007}. 
However, since the chance probability of detecting four low- and high-frequency IP GPs within one $\mu$s is zero, it seems 
likely that the high-frequency IP is related to the low-frequency IP, as seen by \citet{Popov:2008} in previous 
correlations of IP GPs at 600 and 4850 MHz. One possibility is that both the high and low-frequency IP are reflections of 
the MP off of the magnetosphere \citep{Petrova:2009}. In this scenario, radio emission from the MP originating deep in the 
magnetosphere propagates through the electron$-$positron plasma that fills the magnetosphere. Tranverse scattering causes 
the MP emission to be backscattered, causing it to arrive at a different pulse phase. If this was the case, we would expect 
a linear relationship between the strength of the MP and IP in the folded profile for each observation. As shown in Figure 
\ref{fig:mpvsip}, we do see such a relationship at both 1.2 GHz and 330 MHz. The slopes for the 1.2 GHz and 330 MHz 
relations are 0.310$\pm$0.004 and 0.58$\pm$0.08, respectively.

\section{Correlating Radio GPs from the GB43 with $\gamma$-ray Photons from \emph{Fermi}}
\label{fermicorr}

One of the predictions of \citet{Lyutikov:2007} is that there would be increased $\gamma$-ray flux during a GP. In order 
for this model toaccurately reproduce the data, the plasma density of the GP emission region must be $\sim$10$^{5}$ higher 
than the minimum Goldreich-Julian density and the duty cycle of the pulsar must be 0.001. This increase in density could be 
due to enhanced pair production in the pulsar magnetosphere. If correct, these high-energy particles, produced during 
reconnection close to the Y point, where the last closed magnetic field lines approach the light cylinder at the magnetic 
equator \citep{Lyutikov:2007}, are expected to produce curvature radiation from 0.1$-$100 GeV (depending on the value of the 
Lorentz factor $\gamma$) at the time of a GP. This curvature radiation would cause an increase in $\gamma$-rays at the 
times of radio GPs.

Data were downloaded from the \emph{Fermi} online 
archive\footnote{\url{http://fermi.gsfc.nasa.gov/cgi-bin/ssc/LAT/LATDataQuery.cgi}} for days when radio observations 
occured. `Source' class events from Pass 7 data above 100 MeV were selected in an energy dependent radius 
($\theta$ $<$ Max(6.68 $-$ 1.76log(E),1.3)$^\circ$, where E is the energy of the photon in MeV) around the position of 
the pulsar \citep{Abdo:2010}. Only photons in Good Time Intervals (GTIs) were selected and those with a zenith angle $>$ 
100$^\circ$ were excluded to discriminate against $\gamma$-rays generated in the Earth's atmosphere. The photons were 
converted to infinite frequency at the solar system barycenter using the \emph{gtbary} utility from the Fermi Science Tools 
package.

We then searched for coincidences between radio GPs and $\gamma$-ray photons. Due to clock errors in the backend 
on four days at 1.2 GHz (55097, 55257, 55290, 55346), only 75131 MP and 10771 IP GPs were used. However, all of the 5232 MP 
and 888 IP GPs recorded at 330 MHz were used. This resulted in a total of 92022 GPs and 393 $\gamma$-ray photons, with an 
average $\gamma$-ray photon rate of $\sim$16 photons/hour, in agreement with the 15 photons/hour seen by \citet{Abdo:2010}. 
\emph{Fermi} observes the Crab pulsar for about 11 hours per day, resulting in about 34 hours of simultaneous observing 
time. We also made energy cuts and searched for coincidences between radio GPs and $\gamma$-ray photons above both 500 MeV 
and 1 GeV.

\begin{figure}[ht]
\centering
\includegraphics[scale=0.5, angle=-90]{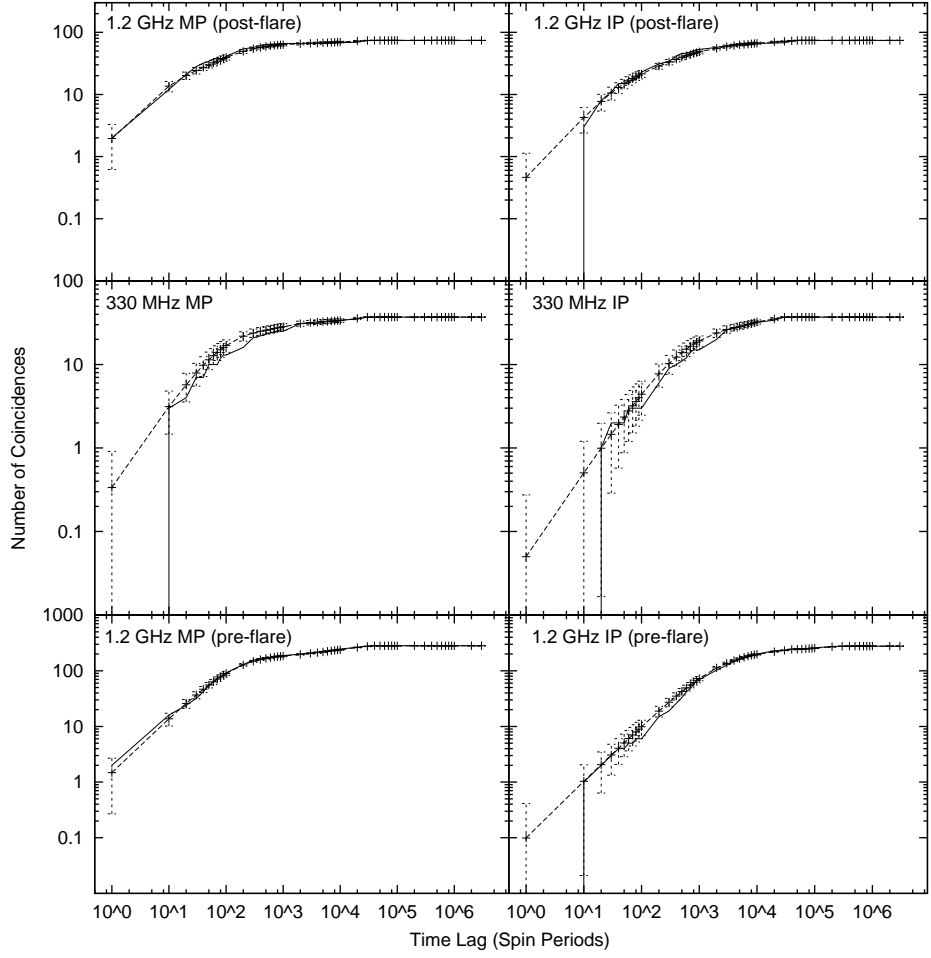}
\caption{Number of coincidences between \emph{Fermi} photons and GB43 GP TOAs (solid lines), as well as the number of 
coincidences between randomized \emph{Fermi} photons and GP TOAs (dotted line) with 1$\sigma$ error bars. The bottom row is 
the correlation with the 1.2 GHz pre-flare data, the middle row is the 330 MHz data, and the top row is the 1.2 GHz 
post-flare data. For each row the MP is the left panel and the IP is the right panel.}
\label{fig:fermi_corr}
\end{figure}

Based on comparisons with correlations of randomized data, we found no significant correlation between MP or IP GPs at both 
1.2 GHz and 330 MHz and $\gamma$-ray photons out to a time lag of $\pm3\times10^6$ spin periods. Since there was not 
necessarily the same number of randomized $\gamma$-ray photons during an observation as there were real $\gamma$-ray 
photons, we used this large time lag to make sure that there were the same number of real and randomized correlations once 
all of the $\gamma$-ray photons were included. The mean and standard 
deviation of the randomized correlations were calculated by randomly assigning $\gamma$-ray photon arrival times within 
GTIs and correlating them with GP TOAs. Figure \ref{fig:fermi_corr} shows the results of this correlation analysis. The 
random correlations are the mean of 10000 trials, and the error bars represent one standard deviation. The largest 
deviation from the mean is a 2.4$\sigma$ anticorrelation between $\gamma$-ray photons and post-flare 1.2 GHz MP GPs at a 
time lag of 20000 spin periods, while the largest correlation is 2.1$\sigma$ for the pre-flare IP at 1.2 GHz at a time lag 
of 200000 spin periods. The maximum correlation/anticorrelation for each frequency at each energy cut is shown in Table 
\ref{table:corrs}. For both the MP and IP for each of the three data sets (pre-flare 1.2 GHz, 330 MHz, post-flare 1.2 GHz), 
there are at most two time lags where the correlation exceeds 2$\sigma$. There are a total of 49 time lags included in the 
correlation, so two time lags are only four percent of the total data, while we statistically expect five percent of the 
data to have correlations beyond 2$\sigma$. Given the low significance and large time lag of these results, they do not 
provide any compelling case for a physical origin.

Selecting only $\gamma$-ray photons above 500 Mev resulted in a total of 119 photons. The maximum correlation for this more 
restricted set was 2.3$\sigma$ at a time lag of one spin period for pre-flare MP at 1.2 GHz. The maximum anticorrelation 
was 3.5$\sigma$ at a time lag of 20000 spin periods for the post-flare MP at 1.2 GHz. No more than one time lag for any 
data set was had a correlation over 2$\sigma$, so those measurements above 2$\sigma$ are statistically insignificant. 
Statistically, only half of a percent of the data should be above 3$\sigma$, and two percent of our data have a 3.5$\sigma$ 
significance. However, this is likely due to the coarseness of our time lags. Given more time lags, we would expect this 
significance to drop below 3$\sigma$.

Only 65 $\gamma$-ray photons had energies above 1 GeV. The maximum correlation occured in the 1.2 GHz pre-flare MP with a 
3.2$\sigma$ significance at a time lag of one spin period. The maximum anticorrelation was 1.2$\sigma$, which occured in 
the 1.2 GHz pre-flare IP at a time lag of 200 spin periods. As with the 500 MeV energy cut, we only see at most one time 
lag with above 2$\sigma$, which is insignificant. Similarly, the 3.2$\sigma$ siginificance is likely due to the coarseness 
of our time lags, and would likely drop given more time lags.

\section{Crab Nebula $\gamma$-ray Flare}
\label{flare}

The recent $\gamma$-ray flare from the Crab Nebula was detected by the AGILE satellite above 100 MeV \citep{Tavani:2011}. 
Elevated $\gamma$-ray flux was observed from MJDs 55457$-$55461. No variations in pulse shape were found at $\gamma$-ray 
\citep{Hays:2010}, X-ray \citep{Tavani:2011}, or radio \citep{Espinoza:2010} energies. \citet{Espinoza:2010} also found no 
increase in pulsed radio flux, glitches, or changes in DM around the date of the flare. Our closest observations before the 
flare were on MJD 55412 at 330 MHz and MJD 55352 at 1.2 GHz, and our first observation after the flare was on MJD 55516 at 
1.2 GHz.

\begin{figure}[ht]
\centering
\includegraphics[scale=0.7, angle=-90]{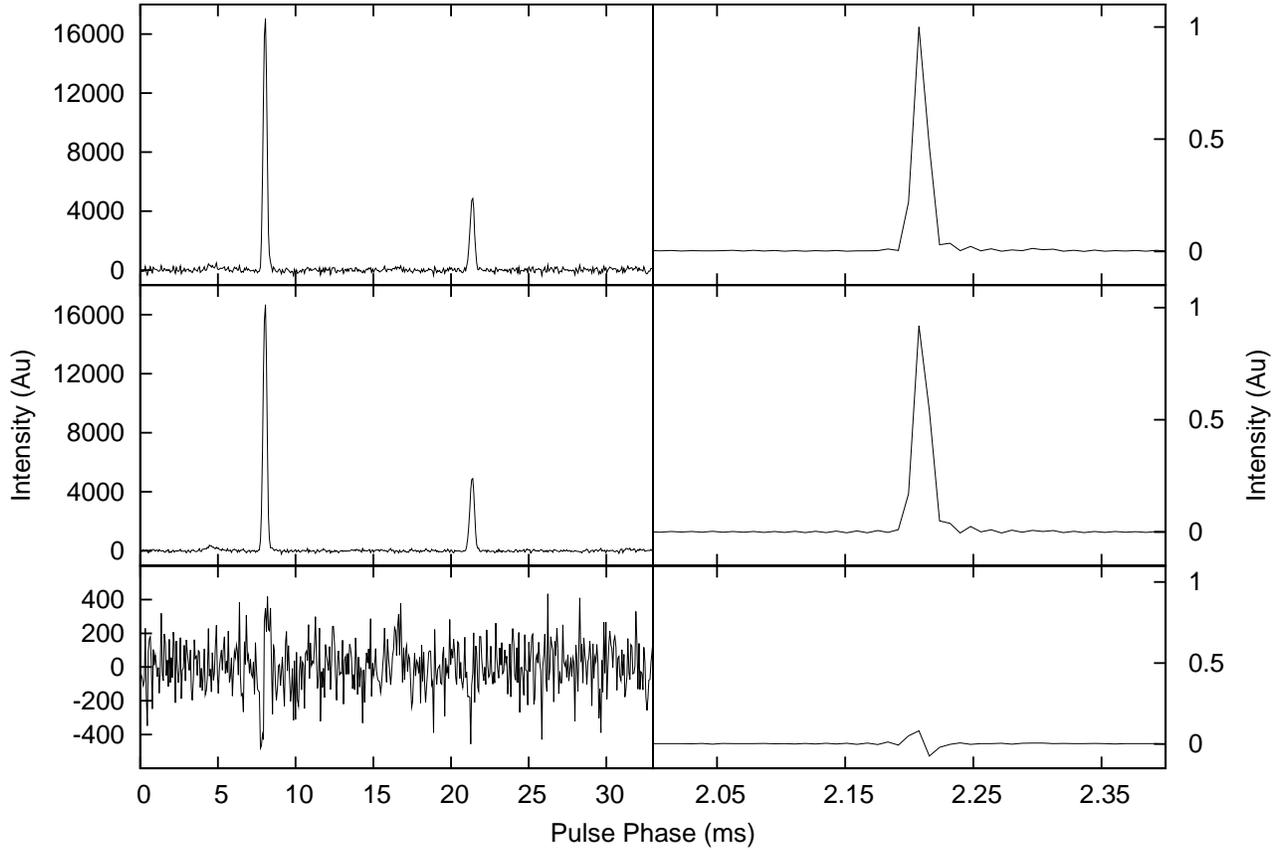}
\caption{A $\gamma$-ray flare occured from MJDs 55457$-$55461, so we checked for changes in the average pulse shape and GP 
shape. {\it Left:} The folded pulse profiles from 1.2 GHz GB43 observations on MJD 55352 (top) and MJD 55532 (middle), and 
the difference between them (bottom). The profiles are scaled so that the area under the main pulse for both profiles is 
the same. As can be seen from the bottom plot, the difference is on the order of the rms noise, so there is no 
statistically significant change in pulse shape. {\it Right:} The average GP profiles from 1.2 GHz GB43 observations on MJD 
55352 (top) and MJD 55532 (middle), and the difference between them (bottom). This difference is not significant as there 
is a larger variation in average GP shape between pre-flare days. The profiles are also scaled so that the area under the 
main pulse for both profiles is the same.}
\label{fig:pulse_diff}
\end{figure}

We looked for changes in the average pulse profile, GP shape, power-law index, and $\gamma$-ray correlation in our pre- and 
post-flare 1.2 GHz data. We found no significant differences in the pulse profile shape and the average GP shape (Figure 
\ref{fig:pulse_diff}) before and on four days about two months after the flare. The pre-flare MP power-law indices are in 
the range 2.1$\pm$0.3 to 3.1$\pm$0.2, while the post-flare power-law indices for the MP and IP are 2.56$\pm$0.05 to 
2.93$\pm$0.05 and 2.4$\pm$0.4 to 2.81$\pm$0.03, respectively. We were unable to calculate power-law indices for the IP 
pre-flare due to low IP GP statistics. The largest anticorrelation between $\gamma$-ray photons and GPs was seen in the 
post-flare data, but it is still on the order of correlations/anticorrelations seen in pre-flare data, both at 1.2 GHz and 
330 MHz.

\section{Conclusions}
\label{conc}

We compared GB43 and GBT GPs and found that 3\% of MP GPs and 5\% of IP GPs at 1.2 GHz were simultaneously detected at 8.9 
GHz. The probability of either of these events occuring by chance is zero. Also, an additional four IP GPs at 1.2 GHz were 
within one $\mu$s of an IP GP at 8.9 GHz, with a zero percent chance probability. This may suggest that, although the 
folded profiles are much different at the two frequencies, the emission mechanism is similar. However, the low percentage 
of low- and high- frequency IP GPs within one $\mu$s could mean that the IP emission mechanism is different at higher 
frequencies, especially since the high-frequency IP is shifted by 10 degrees.

Long observations allowed us to collect the largest sample of GPs to date, which we then used to calculate power-law 
indices for fits to amplitude distributions. These power-law indices agree with previously published values. A 
comparison of these indices to other source classes shows that GP emission is not exactly related to emission from 
other source classes and most closely matches the emission from magnetars and RRATs.

We found no significant correlations between GB43 GPs and \emph{Fermi} $\gamma$-ray photons in the energy range 0.1$-$100 
GeV. There are only a few correlations/anticorrelations in the MP and IP at both frequencies. They occur at different time 
lags and are within 2.5$\sigma$ of the mean for a correlation with randomized data. This suggests that although increased 
pair production in the magnetosphere may contribute to GP occurence, it is not a dominant factor. More likely possibilities 
for GP generation are increased coherence or changes in beaming.

Multifrequency correlations have been searched for previously. \citet{Shearer:2003} found a 3\% increase in the brightness 
of optical pulses at the time of GPs and \citet{Collins:2012} found a slight correlation between GPs and enhanced optical 
pulses, which supports our $\sim$2$\sigma$ correlations/anticorrelations in suggesting that there are small fluctuations in 
the magnetospheric particle density during GPs. The anticorrelations, however, would suggest that increased particle 
density would lead to more radio emission and less $\gamma$-ray emission, which does not make sense in the context of 
Lyutikov's theory. Since the significance of the anticorrelations are on the order of the significance of the correlations, 
we assume that these are insignificant. \citet{Lundgren:1995} found that the $\gamma$-ray flux does not vary by more than 
2.5 times the average flux during a GP. \citet{Bilous:2011} ruled out a strong correlation between GPs at 8.9 GHz and 
$\gamma$-ray photons above 100 MeV. They were able to put an upper limit on the $\gamma$-ray flux during IP GPs of 8$-$16 
times the average pulsed flux, suggesting that there still might be a slight correlation between GPs and $\gamma$-ray 
photons. We could not carry out the same experiment because we did not have sufficient $\gamma$-ray photons to compare the 
$\gamma$-ray profile made with $\gamma$-ray photons around GPs with the $\gamma$-ray profile excluding $\gamma$-ray photons 
around GPs.

The recent $\gamma$-ray flare from the Crab Nebula occured during the span of our observations, so we were able to compare 
the behavior of the Crab pulsar before and after the flare. We found no significant changes in pulse shape, power-law 
index, or $\gamma$-ray correlation, suggesting that there was no change in the properties of the pulsar during the 
flare.

\section{Acknowledgements}

The National Radio Astronomy Observatory is a facility of the National Science Foundation operated under cooperative 
agreement by Associated Universities, Inc. This work was supported by NASA \emph{Fermi} Grant NNX10AD14G and a WV EPSCoR 
grant.

\clearpage

\begin{table}[ht]
\centering
\caption{Observational parameters including the MJD the observation was taken, the center frequency used, the length of the 
observation, the sampling time, the DM used for dedispersion, and the number of GPs found with S/N $\ge$ 10. The starred 
MJDs are days in which simultaneous observations with the GBT occured.}
\begin{tabular}{c c c c c c c}
\\
\hline
\hline
MJD & Frequency	& Observation &	Sampling & DM &	Number & Number \\
 & (MHz) & Length (min) & Time ($\mu$s) & (pc cm$^{-3}$)  & of MP GPs & of IP GPs \\ \hline
55079 & 1200 & 5 & 204.8 & 56.8005 & 33 & 1 \\
55079 & 1200 & 108 & 204.8 & 56.8005 & 341 & 27 \\
*55083 & 1200 & 4 & 204.8 & 56.8005 & 19 & 1 \\
*55083 & 1200 & 120 & 204.8 & 56.8005 & 582 & 42 \\
55085 & 1200 & 120 & 204.8 & 56.8005 & 73 & 0 \\
*55086 & 1200 & 156 & 204.8 & 56.8005 & 1677 & 96 \\
*55090 & 1200 & 150 & 204.8 & 56.8005 & 551 & 31 \\
*55093 & 1200 & 168 & 204.8 & 56.8005 & 650 & 42 \\
*55095 & 1200 & 108 & 51.2 & 56.8005 & 669 & 47 \\
*55096 & 1200 & 150 & 204.8 & 56.8005 & 264 & 21 \\
*55097 & 1200 & 240 & 204.8 & 56.8005 & 519 & 45 \\
*55099 & 1200 & 210 & 204.8 & 56.8005 & 1435 & 76 \\
*55102 & 1200 & 150 & 204.8 & 56.8005 & 1228 & 76 \\
55118 & 1200 & 6 & 204.8 & 56.8109 & 2 & 0 \\
55118 & 1200 & 6 & 51.2 & 56.8109 & 16 & 0 \\
55139 & 1200 & 240 & 51.2 & 56.8229 & 3757 & 308 \\
55142 & 1200 & 120 & 51.2 & 56.8229 & 244 & 22 \\
55172 & 1200 & 22 & 51.2 & 56.8279 & 25 & 0 \\
55178 & 1200 & 282 & 51.2 & 56.8279 & 1808 & 110 \\
55180 & 1200 & 114 & 51.2 & 56.8279 & 478 & 27 \\
55240 & 1200 & 7 & 245.76 & 56.8053 & 32 & 2 \\
55240 & 1200 & 180 & 122.88 & 56.8053 & 1575 & 101 \\
55256 & 1200 & 15 & 61.44 & 56.8622 & 148 & 8 \\
\hline
\end{tabular}
\label{table:crab_data}
\end{table}

\clearpage

\begin{table}[ht]
\centering
\begin{tabular}{c c c c c c c}
\\
\hline
\hline
MJD & Frequency & Observation & Sampling & DM & Number & Number \\
 & (MHz) & Length (min) & Time ($\mu$s) & (pc cm$^{-3}$)  & of MP GPs & of IP GPs\\ \hline
55257 & 1200 & 198 & 245.76 & 56.8622 & 1098 & 76 \\
55261 & 1200 & 492 & 245.76 & 56.8622 & 41 & 9 \\
55264 & 1200 & 15 & 61.44 & 56.8622 & 106 & 5 \\
55269 & 1200 & 360 & 245.76 & 56.8622 & 42 & 2 \\
55270 & 1200 & 30 & 245.76 & 56.8622 & 291 & 17 \\
55284 & 1200 & 180 & 245.76 & 56.8622 & 1094 & 63 \\
55290 & 1200 & 108 & 245.76 & 56.8228 & 169 & 10 \\
55299 & 1200 & 10 & 245.76 & 56.8228 & 4 & 0 \\
55304 & 1200 & 60 & 245.76 & 56.8228 & 73 & 5 \\
55346 & 1200 & 60 & 245.76 & 56.8022 & 672 & 40 \\
55347 & 1200 & 120 & 245.76 & 56.8022 & 1170 & 76 \\
55347 & 1200 & 168 & 61.44 & 56.8022 & 2701 & 243 \\
55352 & 1200 & 240 & 61.44 & 56.8022 & 2182 & 146 \\
55403 & 330 & 37 & 223.42 & 56.7988 & 271 & 31 \\
55405 & 330 & 60 & 819.2 & 56.7988 & 1201 & 250 \\
55406 & 330 & 222 & 819.2 & 56.7988 & 1775 & 273 \\
55411 & 330 & 60 & 819.2 & 56.7962 & 1332 & 220 \\
55412 & 330 & 30 & 819.2 & 56.7962 & 653 & 114 \\
55516 & 1200 & 550 & 245.76 & 56.8065 & 4870 & 326 \\
55532 & 1200 & 315 & 61.44 & 56.7964 & 11671 & 937 \\
55539 & 1200 & 10 & 61.44 & 56.7964 & 786 & 146 \\
55541 & 1200 & 335 & 61.44 & 56.7964 & 33611 & 7687 \\
\hline
\end{tabular}
\end{table}

\clearpage

\begin{table}[ht]
\centering
\caption{Sampling times, DM smearing times, scattering times, and effective resolution at the top and bottom of the band 
for the GB43 330 MHz and 1.2 GHz observations, as well as for the GBT 350 MHz observation. The DM smearing times were 
calculated using the delay between two consecutive channels caused by dispersion, and the scattering times were estimated 
from \citet{Kuzmin:2011}. The sampling time listed for the 1.2 GHz GB43 observation was the most common sampling time from 
Table \ref{table:crab_data}.}
\begin{tabular}{c c c c c c c c}
\\
\hline
\hline
Frequency & Sampling Time & \multicolumn{2}{c}{DM Smearing Time} & \multicolumn{2}{c}{Scattering Time} & \multicolumn{2}{c}{Effective Resolution} \\
(MHz) & (us) & Top & Bottom & Top & Bottom & Top & Bottom \\
\hline
330 (GB43) & 819.2 & 300~$\mu$s & 2.4~ms & 470~$\mu$s & 5.8~ms & 990~$\mu$s & 6.3~ms \\
350 (GBT) & 81.92 & 361~$\mu$s & 855~$\mu$s & 670~$\mu$s & 1.9~ms & 765~$\mu$s & 2.0~ms \\
1200 (GB43) & 204.8 & 23~$\mu$s & 180~$\mu$s & 4~$\mu$s & 54~$\mu$s & 206~$\mu$s & 278~$\mu$s \\
\hline
\end{tabular}
\label{table:times}
\end{table}

\clearpage

\begin{table}[ht]
\centering
\caption{Differential power-law measurements for the MP and IP for each observing epoch. Some MJDs have two entries
listed because there were two separate observations on those days. An X means that there were not enough pulses recorded
for the MP or IP on that day to fit a power-law to the amplitude distribution.}
\begin{tabular}{c c c}
\\
\hline
\hline
MJD & Differential Power- & Differential Power- \\
 & law Index (MP)& law Index (IP) \\
\hline
55079 & X & X \\
55079 & 2.1$\pm$0.3 & X \\
55083 & X & X \\
55083 & 3.1$\pm$0.2 & X \\
55085 & X & X \\
55086 & 2.70$\pm$0.06 & X \\
55090 & 2.8$\pm$0.1 & X \\
55093 & 2.6$\pm$0.2 & X \\
55095 & 2.9$\pm$0.2 & X \\
55096 & X & X \\
55097 & X & X \\
55099 & 2.93$\pm$0.07 & X \\
55102 & 2.92$\pm$0.08 & X \\
55118 & X & X \\
55118 & X & X \\
55139 & 2.61$\pm$0.08 & X \\
55142 & X & X \\
55172 & X & X \\
55178 & 2.81$\pm$0.08 & X \\
55180 & X & X \\
55240 & X & X \\
55240 & 2.1$\pm$0.1 & X \\
55256 & X & X \\
\hline
\end{tabular}
\label{table:indices}
\end{table}

\clearpage

\begin{table}[ht]
\centering
\begin{tabular}{c c c}
\\
\hline
\hline
MJD & Differential Power- & Differential Power- \\
 & law Index (MP) & law Index (IP) \\
\hline
55257 & 2.1$\pm$0.1 & X \\
55261 & X & X \\
55264 & X & X \\
55269 & X & X \\
55270 & X & X \\
55284 & 2.1$\pm$0.1 & X \\
55290 & X & X \\
55299 & X & X \\
55304 & X & X \\
55346 & 2.4$\pm$0.2 & X \\
55347 & 2.8$\pm$0.1 & X \\
55347 & 2.40$\pm$0.07 & X \\
55352 & 2.21$\pm$0.07 & X \\
55403 & 2.5$\pm$0.2 & X \\
55405 & 2.95$\pm$0.09 & 2.4$\pm$0.2 \\
55406 & 2.94$\pm$0.05 & 3.1$\pm$0.2 \\
55411 & 2.8$\pm$0.1 & 2.9$\pm$0.2 \\
55412 & 2.7$\pm$0.2 & X \\
55516 & 2.93$\pm$0.05 & 2.4$\pm$0.4 \\
55532 & 2.56$\pm$0.05 & 2.7$\pm$0.2 \\
55541 & 2.8$\pm$0.1 & 2.81$\pm$0.03 \\
\hline
\end{tabular}
\end{table}

\clearpage

\begin{table}[ht]
\centering
\caption{Comparison of differential power-law indices for the MP and IP at both 1.2 GHz and 330 MHz between this work and
previously published values.}
\begin{threeparttable}
\begin{tabular}{c c c c}
\\
\hline
\hline
Frequency & Differential Power- & Differential Power- & Reference \\
(MHz) & law Index (MP) & law Index (IP) & \\
\hline
112 & 3.3\tnote{a} & $-$ & \citet{Smirnova:2009} \\
146 & 3.5 & 3.8 & \citet{Argyle:1972} \\
200 & 2.7\tnote{a} & $-$ & \citet{Bhat:2007} \\
330 & 2.5$-$3.0 & 2.4$-$3.1 & This work \\
430 & 2.3 & $-$\tnote{b} & \citet{Cordes:2004} \\
600 & 3.2 & 3.0 & \citet{Popov:2009} \\
812 & 3.3\tnote{a} & $-$ & \citet{Lundgren:1995} \\
1200 & 2.7$-$4.2 & 2.6 & \citet{Popov:2007} \\
1200 & 2.1$-$3.1 & 2.4$-$2.8 & This work \\
1300 & 2.3\tnote{a} & $-$ & \citet{Bhat:2008} \\
1400 & 2.8 & 3.1 & \citet{Karuppusamy:2010} \\
2100 & 3.0\tnote{a} & $-$ & \citet{Zhuravlev:2011} \\
4850 & 2.8\tnote{a} & $-$ & \citet{Popov:2008} \\
\hline
\end{tabular}
\begin{tablenotes}
\item[a] MP and IP GPs were combined in these analyses
\item[b] No measurement was taken for the IP
\end{tablenotes}
\end{threeparttable}
\label{table:indices_comp}
\end{table}

\clearpage

\begin{table}[ht]
\centering
\caption{Differential power-law indices for different source classes for comparison with GP power-laws.}
\begin{tabular}{c c c}
\\
\hline
\hline
Source & Power-law & Reference \\
Class & Index & \\
\hline
Normal Pulsars & 3.85 & \citet{Kramer:2002} \\
Magnetars & 2.1$-$7.7 & \citet{Serylak:2009} \\
RRATs & 3.0 & \citet{Miller:2012} \\
\hline
\end{tabular}
\label{table:source_class}
\end{table}

\clearpage

\begin{sidewaystable}[ht]
\centering
\caption{Maximum correlation/anticorrelation for each frequency at each energy cut, and the time lags at which they occur. 
Note that the maximum correlation for the 330 MHz IP with the 500 MeV energy cut is negative.}
\begin{tabular}{c c c c c c c}
\\
\hline
\hline
Energy & Frequency & MP/IP & Max & Time Lag & Max Anti- & Time Lag \\
 Cut & (MHz) & & Correlation ($\sigma$) & (Spin Periods) & correlation ($\sigma$) & (Spin Periods) \\
\hline
\multirow{6}{*}{100 MeV} & \multirow{2}{*}{Pre-flare 1200} & MP & +1.4 & 30000 & $-$0.9 & 30 \\
 & & IP &+2.1 & 200000 & $-$1.7 & 300 \\
 & \multirow{2}{*}{330} & MP & +0.2 & 10000 & $-$2.1 & 200 \\
 & & IP & +0.5 & 30 & $-$1.5 & 1000 \\
 & \multirow{2}{*}{Post-flare 1200} & MP & +1.5 & 40 & $-$2.4 & 20000 \\
 & & IP & +1.9 & 500 & $-$2.3 & 30000 \\
\hline
\multirow{6}{*}{500 MeV} & \multirow{2}{*}{Pre-flare 1200} & MP & +2.3 & 1 & $-$0.4 & 8000 \\
 & & IP & +1.5 & 20000 & $-$1.5 & 700 \\
 & \multirow{2}{*}{330} & MP & +0.1 & 6000 & $-$1.7 & 200 \\
 & & IP & $-$0.1 & 300 & $-$1.7 & 2000 \\
 & \multirow{2}{*}{Post-flare 1200} & MP & +0.8 & 200 & $-$3.5 & 20000 \\
 & & IP & +1.0 & 200 & $-$2.7 & 30000 \\
\hline
\multirow{6}{*}{1 GeV} & \multirow{2}{*}{Pre-flare 1200} & MP & +3.2 & 1 & $-$0.1 & 60 \\
 & & IP & +2.0 & 5000 & $-$1.2 & 200 \\
 & \multirow{2}{*}{330} & MP & +0.8 & 6000 & $-$0.9 & 10 \\
 & & IP & +0.9 & 6000 & $-$1.1 & 100 \\
 & \multirow{2}{*}{Post-flare 1200} & MP & +1.9 & 60 & $-$0.4 & 1 \\
 &  & IP & +2.1 & 3000 & $-$0.7 & 10 \\
\hline
\end{tabular}
\label{table:corrs}
\end{sidewaystable}

\end{document}